\documentstyle[12pt]{article}
\newlength{\dinwidth}
\newlength{\dinmargin}
\begin{document}
\def\thefootnote{\fnsymbol{footnote}}
\thispagestyle{empty}
\begin{flushright}
\begin{tabular}{l}
FTUAM-92/12\\\vspace*{24pt}
June, 1992
\end{tabular}
\end{flushright}

\vspace*{1.5cm}

{\vbox{\centerline{{\Large{\bf DUALITY IN NON-TRIVIALLY COMPACTIFIED
}}}}}
\vskip12pt\centerline{{\Large{\bf HETEROTIC STRINGS
}}}

\vskip72pt\centerline{M.A.R. Osorio\footnote{Bitnet address:
OSORIO@EMDUAM11} and M. A. V\'azquez-Mozo\footnote{Bitnet address:
MAVAZ@EMDUAM11}}

\vskip12pt
\centerline{{\it Departamento de F\'{\i}sica Te\'orica C-XI}}\vskip2pt
\centerline{{\it Universidad Aut\'onoma de Madrid}}\vskip2pt
\centerline{{\it 28049 Madrid, Spain}}

\vskip .7in

\indent

We study the implications of duality symmetry on the analyticity
properties of the partition function as it depends upon the
compactification length.  In order to obtain non-trivial
compactifications, we give a physical prescription to get the Helmholtz
free energy for any heterotic string supersymmetric or not.  After
proving that the free energy is always invariant under the duality
transformation $R\rightarrow \alpha^{'}/(4R)$ and getting the
zero temperature theory whose partition function
corresponds to the Helmholtz potential, we show that the self-dual point
$R_{0}=\sqrt{\alpha^{'}}/2$ is a generic singularity as the Hagedorn
one.  The main difference between these two critical compactification
radii is that the term producing the singularity at the self-dual point
is finite for any $R \neq R_{0}$.  We see that this behavior at $R_{0}$
actually implies a loss of degrees of freedom below that point.

\setcounter{page}{0}
\newpage

\section{Introduction}

Recently some thinking effort has been devoted to the problem of
$R-$duality in string theory \cite{EalvaMar2}\cite{GrossKlebanov}.
One of
the related topics that seems to have received less attention is that of
the implications of duality on the analytic structure of the partition
function, $Z(R)$, as a function of the compactification radius $R$. As
far as we know this problem has only been fully
treated  for the non-critical string coupled to conformal matter
with $c=1$ in which $Z(R)$ is $C^{\infty}({\bf R^{+}})$ as a function
of $R$
\cite{GrossKlebanov}\cite{c=1}.
In the case of critical heterotic strings this problem seems to have
been studied only in \cite{KutasovSeiberg}\cite{MarMavaz}. In these
works
a small class of heterotic strings has been treated: trivial toroidal
compactifications on ${\bf R}^{(critical\hspace{2mm}d)-1} \times
S^{1}$. On the other hand it is also well known that there exists a
completely analogous duality invariance (the so called $\beta-$duality
in opposition to space-time duality)
for the thermal free energies
corresponding to the family of supersymmetric heterotic strings. This is
the only case in which duality is a mathematically well defined property
of the
Helmholtz free energy although there is always a well defined duality
relationship for the integrand of the free energy represented as an
integral over the fundamental region of the modular group, even for the
bosonic string. Of course duality is actually an invariance of the
spectrum of the theory

Using the non-trivial relationship between the Helmholtz free energy and
the partition function of the same theory on ${\bf
R}^{(critical\hspace{2mm}
d)-1} \times S^{1}$ we can get an enormous class of
non-supersymmetric strings which exhibit non-trivial duality as a
property of their corresponding partition functions. By non-trivial
duality we mean that under the transformation
$R \rightarrow constant/R$ the solitonic contributions associated with
each spin structure interchange between them.  In other words, there
is a correlation between the solitonic contributions and the spin
structures. The study of heterotic
strings at finite temperature is then of interest to understand
spacetime duality as a physical property.

On the other hand the study of superstrings at finite temperature has
interest on its own. A lot of effort
has been devoted to this topic (cf. for example
\cite{Ealva}\cite{Polchinski}\cite{Tan}), without, in our opinion,
getting any conclusive answer for the most important questions: is there
any possibility of a phase transition at the Hagedorn temperature (or
before or after)? And
if there is any, what would the number of physical degrees of freedom be
at high temperature?
The number of degrees of freedom of the Nambu-Goto string is also a
piece of
major interest in order to know whether this string has something to do
with the features of the deconfinement phase of QCD (cf. \cite{Polchi}).

The existence of
$\beta-$duality for the free energy is a puzzling property of
the heterotic
string closely related to the latter question. Namely, at one loop
level $\beta-$duality on the Helmholtz free energy reads
\begin{equation}
F_{het}(\beta)=\frac{\pi^2}{\beta^2}
F_{het}\left(\frac{\pi^2}{\beta}\right)
\label{duality}
\end{equation}
The presence of the Hagedorn length $\beta_{H}$ together with the
$\beta-$duality property implies that there exists another
critical length
$\beta_{H}^{*}=\pi^{2}/\beta_{H}$
such that although $F_{H}(\beta)$ diverges for
$\beta_{H}^{*}<\beta<\beta_{H}$ it is finite for $\beta<\beta_{H}^{*}$.
Using $F_{het}(\beta)$ when $\beta<\beta_{H}^{*}$ as the thermodynamical
Helmholtz potencial some shocking thermodynamical features appear for
this would-be high temperature phase.
The most striking one is that
in the limit $\beta \rightarrow 0^{+}$, (\ref{duality}) implies that the
free energy behaves as
\begin{equation}
F_{het}(\beta) \sim \frac{\pi^2}{\beta^2} \Lambda
\label{limit}
\end{equation}
Here $\Lambda$ is the cosmological constant; since $\Lambda=0$ for a
heterotic supersymmetric string (\ref{limit}) implies that no
degree of freedom will survive at high temperature. In fact we will show
that (\ref{duality}) and consequently (\ref{limit}) also hold for any
heterotic string,
even a non-supersymmetric one.
Some authors \cite{AtickWitten} (cf. also \cite{Polchi}) have pointed
out that this could indicate that the heterotic string
would be described by a topological theory in that limit.
Thermodynamically things appear as though
Bose and Fermi statistics were not equivalent at high temperature.

In the present work we will carefully study the
behaviour of thermal heterotic strings
to prove, among other things, that
above the Hagedorn length, which is the same for every heterotic
string,
there is one more critical length at the self-dual point  which
is generic
too and does not correspond to a divergent term in the free energy.

In section $2$ the prescription given by Atick and Witten in
\cite{AtickWitten} for
constructing the free energy of the heterotic string at one loop
will be recovered by using what we regard as a much more
physical prescription.
In fact, by making use of the results of reference
\cite{EalvaMar2} we will be able to show that every heterotic string
enjoys $\beta-$duality, even in the non-supersymmetric case. With this
new prescription we will explicitly
calculate the free energy  for the
family of non-supersymmetric heterotic strings in two dimensions
which appear in \cite{Moore}\cite{GinspargVafa}\cite{MarMavaz}.

Section $3$ will be devoted to the
study of the possibility
of getting separately the free
energy for the bosonic and fermionic sectors of the heterotic
string in a manifestly modular invariant way.
This goal is a legitimate one from a purely
thermodynamical point of view (e.g. if we are interested in studying a
possible Bose condensation). The results
so obtained will be used to analyze what would be the structure of the
high temperature phase (or equivalently the
number of degrees of freedom at high energy).

In section $4$
we shall study the behaviour of the free energy at some given values of
$\beta$ at which special generic singularities appear
\cite{KutasovSeiberg}\cite{MarMavaz}. These singularities will
be interpreted as coming from contributions of states which actually
behave as ghosts, killing the physical degrees of freedom that are a
surplus for duality to hold. Finally in section $5$ we will summarize
the conclusions.

\section{Getting the free energy}

In reference \cite{AtickWitten} Atick and Witten gave a prescription for
computing the free energy for the heterotic string. In their approach
the contribution of an arbitrary set of windings $(n,m)$ is weighted, at
one loop level, by a phase given by
\begin{eqnarray}
U_{1}(n,m)&=&\frac{1}{2}\left(-1+(-1)^{n}+(-1)^{m}+(-1)^{m+n}\right)
\nonumber \\
U_{2}(n,m)&=&\frac{1}{2}\left(1-(-1)^{n}+(-1)^{m}+(-1)^{m+n}\right)
\\
U_{3}(n,m)&=&\frac{1}{2}\left(1+(-1)^{n}+(-1)^{m}-(-1)^{m+n}\right)
\nonumber \\
U_{4}(n,m)&=&\frac{1}{2}\left(1+(-1)^{n}-(-1)^{m}+(-1)^{m+n}\right)
\nonumber
\label{phases}
\end{eqnarray}
where $1,2,3,4$ label respectively the four spin structures
$(+,+),(+,-),(-,-),(-,+)$.

Generalizing these phases to arbitrary genus, one can represent the
genus-$g$ contribution to the
free energy per unit volume for the heterotic string in a very
manageable form, namely \cite{EalvaMar2}
\begin{equation}
F_{g}(\beta)=\int_{{\cal F}_{g}}d\mu(m)\sum_{s}
\Lambda_{s}(\tau,\bar{\tau}) \theta \left[
\begin{array}{cc}
{\bf 0} & {\bf 0} \\
{\bf s}_{2} & {\bf s}_{1}
\end{array} \right] (0|{\bf \tilde{\Omega}}_{g})
\label{Freenergy}
\end{equation}
where the cosmological constant is given by
\begin{equation}
\Lambda_{s}=\int_{{\cal
F}_{g}}d\mu(m)\sum_{s}\Lambda_{s}(\tau,\bar{\tau})
\end{equation}
and
\begin{equation}
s \equiv \left[
\begin{array}{c}
{\bf s}_{1} \\
{\bf s}_{2}
\end{array} \right]
\end{equation}
${\bf s}_{1},{\bf s}_{2} \in \left[({\bf Z}/2)/{\bf Z}\right]^{g}$
being
the characteristics defining the $2^{2g}$ spin structures on the Riemann
surface (in fact, only the even spin structures contribute). The
second argument of the Riemann theta function \cite{Mumford} is
\begin{equation}
{\bf \tilde{\Omega}}_{g}={\bf \Omega}_{g}+\frac{1}{2}\left[
\begin{array}{cc}
0 & I \\
I & 0
\end{array}
\right]=
\frac{i\beta^{2}}{2\pi^{2}}\left[
\begin{array}{cc}
\tau_{1}\tau_{2}^{-1}\tau_{1}+\tau_{2} & -\tau_{1}\tau_{2}^{-1} \\
-\tau_{2}^{-1}\tau_{1} & \tau_{2}^{-1}
\end{array}
\right]+\frac{1}{2}\left[
\begin{array}{cc}
0 & I \\
I & 0
\end{array}
\right]
\end{equation}
Manipulating (\ref{Freenergy}) one is able to rewrite
the genus$-g$ contribution to the free energy as
\begin{equation}
F_{g}(\beta)=\int_{{\cal F}_{g}} d\mu(m) \sum_{s}\Lambda_{s} \sum_{t}
(-1)^{4({\bf s}_{2}{\bf t}_{1}+{\bf s}_{1}{\bf t}_{2}+ {\bf t}_{1}{\bf
t}_{2})} \theta \left[
\begin{array}{cc}
{\bf t}_{1} & {\bf t}_{2}\\
0   &   0
\end{array}
\right] (0|4{\bf \Omega}_{g})
\label{upper}
\end{equation}
In \cite{EalvaMar2} it is shown that given this form
we have the following duality
relation for $F_{g}(\beta)$
\begin{equation}
F_{g}(\beta)=\left(\frac{\pi^{2}}{\beta^{2}}\right)^{g}
F_{g}\left(\frac{\pi^{2}}{\beta}\right)
\end{equation}
In particular, for $g=1$ we get (\ref{duality}).

So every heterotic string such that its free energy is
given by the application of the Atick and Witten prescription obeys
duality at arbitrary genus (provided only even spin structures
contribute to the  free energy).

Let us now consider the general form of the one loop free energy for a
heterotic
string. By using the four $SO(10)$ conjugacy classes we can write the
part of the integrand of the cosmological constant which multiplies
Poincar\'e's invariant measure as (cf. for example
\cite{NairShapereWil})
\begin{eqnarray}
\sum_{s} \Lambda_{s}(\tau,\bar{\tau})=
\tau^{-\frac{d-2}{2}}_{2}\left[{\frac{\bar{\theta}_{2}^{4}}
{\bar{\eta}^{12}}z_{s}(\tau,\bar{\tau})+
\frac{\bar{\theta}_{2}^{4}}{\bar{\eta}^{12}}
z_{c}(\tau,\bar{\tau})-
\frac{\bar{\theta}_{3}^{4}-\bar{\theta}_{4}^{4}}
{\bar{\eta}^{12}} z_{v}(\tau,\bar{\tau})-
\frac{\bar{\theta}_{3}^{4}+\bar{\theta}_{4}^{4}}
{\bar{\eta}^{12}} z_{o}(\tau,\bar{\tau}})\right]=
\nonumber
\\
\tau_{2}^{-\frac{d-2}{2}}\left[\frac{\bar{\theta}_{2}^{4}}
{\bar{\eta}^{12}}(z_{s}+z_{c})+
\frac{\bar{\theta}_{4}^{4}}{\bar{\eta}^{12}}(z_{v}-z_{o})-
\frac{\bar{\theta}_{3}^{4}}{\bar{\eta}^{12}}(z_{v}+z_{o})\right]
\label{cosmconst2}
\end{eqnarray}
Where $z_{o},z_{v},z_{s},z_{c}$ are respectively the contributions
associated with the four conjugacy classes of $SO(10)$ (scalar,
vectorial
and two spinorials), and $d$ is the number of non-compact dimensions.
For example, for the supersymmetric heterotic string in ten dimensions
we have that $z_{o}=z_{c}=0$ and $z_{v}=z_{s}=\Theta_{\Gamma_{8} \oplus
\Gamma_{8}}\eta^{-24}$.

{}From the modular invariance of (\ref{cosmconst2}) we get
\begin{eqnarray}
\lefteqn{\frac{\bar{\theta}_{2}^{4}}{\bar{\eta}^{12}}(z_{s}+z_{c})+
\frac{\bar{\theta}_{4}^{4}}{\bar{\eta}^{12}}(z_{v}-z_{o})-
\frac{\bar{\theta}_{3}^{4}}{\bar{\eta}^{12}}(z_{v}+z_{o})=}   \nonumber
\\
& & \frac{\bar{\theta}_{2}^{4}}{\bar{\eta}^{12}}
\left(T(z_{s})+T(z_{c})\right)-
\frac{\bar{\theta}_{3}^{4}}{\bar{\eta}^{12}}
\left(T(z_{v})-T(z_{o})\right)+
\frac{\bar{\theta}_{4}^{4}}{\bar{\eta}^{12}}
\left(T(z_{v})+T(z_{o})\right)
\label{t-trans}
\end{eqnarray}
and
\begin{eqnarray}
\lefteqn{\frac{\bar{\theta}_{2}^{4}}{\bar{\eta}^{12}}(z_{s}+z_{c})+
\frac{\bar{\theta}_{4}^{4}}{\bar{\eta}^{12}}(z_{v}-z_{o})-
\frac{\bar{\theta}_{3}^{4}}{\bar{\eta}^{12}}(z_{v}+z_{o})=
\frac{\bar{\theta}_{4}^{4}}{\bar{\eta}^{12}}
\bar{\tau}^{-4}|{\tau}|^{d-2}
\left(S(z_{s})+S(z_{c})\right)+} \nonumber \\
& & \frac{\bar{\theta}_{2}^{4}}{\bar{\eta}^{12}}\bar{\tau}^{-4}
|{\tau}|^{d-2} \left(S(z_{v})-S(z_{o})\right)-
\frac{\bar{\theta}_{3}^{4}}{\bar{\eta}^{12}}\bar{\tau}^{-4}
|{\tau}|^{d-2} \left(S(z_{v})+S(z_{o})\right)
\label{s-trans}
\end{eqnarray}
where $T$ and $S$ are the two transformations generating the modular
group: $T:\tau \longrightarrow \tau+1$ and  $S:\tau \longrightarrow
-1/\tau$.

These relations determine the transformation properties under both
$T$ and $S$ of the
combination of $z_{i}$'s which appear in (\ref{cosmconst2}).
However we should be very careful about the
identification of the terms multiplying Jacobi's theta function in
(\ref{t-trans}) and (\ref{s-trans}). If we consider a linear combination
of the form
\begin{equation}
A_{1}\bar{\theta}_{2}^{4}+A_{2}\bar{\theta}_{3}^{4}+
A_{3}\bar{\theta}_{4}^{4}=0
\end{equation}
then there exists a trivial solution which correspond to
$A_{1}=A_{2}=A_{3}=0$. But since Jacobi theta functions satisfy the
well known identity
\begin{equation}
\bar{\theta}_{2}^{4}-\bar{\theta}_{3}^{4}+\bar{\theta}_{4}^{4}=0
\end{equation}
we also have a non-trivial solution, namely
\begin{equation}
A_{1}=-A_{2}=A_{3}
\end{equation}
{}From a physical point of view the non-trivial solution would correspond
to the fact that when performing a modular transformation we are not
recovering the same theory we started with. We need to add to it the
spectrum of a supersymmetric heterotic string. Our starting point is
that (\ref{cosmconst2}) corresponds to the complete theory.
Actually if we add a supersymmetric heterotic spectrum we can always
write the resulting cosmological constant as in (\ref{cosmconst2}).
Thus we will only consider the trivial solution of equations
(\ref{t-trans}) and (\ref{s-trans}). This gives the following
transformation properties (cf. for example
\cite{NarainSarmadi}\cite{NairShapereWil})
\begin{eqnarray}
T(z_{s})+T(z_{c})=z_{s}+z_{c}     \nonumber
\\
T(z_{v})+T(z_{o})=z_{v}-z_{o}
\\
T(z_{v})-T(z_{o})=z_{v}+z_{o}     \nonumber
\label{T}
\end{eqnarray}
for the $T$ transformation, and
\begin{eqnarray}
S(z_{s})+S(z_{c})={\bar \tau}^{4}|\tau|^{2-d}(z_{v}-z_{o})    \nonumber
\\
S(z_{v})+S(z_{o})={\bar \tau}^{4}|\tau|^{2-d}(z_{v}+z_{o})
\\
S(z_{v})-S(z_{o})={\bar \tau}^{4}|\tau|^{2-d}(z_{s}+z_{c})    \nonumber
\label{S}
\end{eqnarray}
for $S$. These tranformation properties will be of major interest when
dealing with the theory at finite temperature.

To construct the corresponding free energy one can use an
analog model.
The free energy of a quantum field with $N_{B}$ bosonic physical degrees
of freedom is given at one loop by \cite{Polchinski}\cite{EalvaMar3}
\begin{equation}
F_{B}(\beta)=-N_{B}\int_{0}^{\infty}ds s^{-1-\frac{d}{2}}
\theta_{3}^{'}\left(0\left|\frac{i\beta^{2}}{2\pi s}\right.\right)
e^{-m^{2}s/2}=:N_{B}f_{B}(\beta)
\end{equation}
where $f_{B}(\beta)$ is the free energy per physical degree of freedom
and the prime indicates that the zero mode has been suppressed.
For a fermionic field of the same mass and $N_{F}$ physical degrees of
freedom we have \cite{Osorio}
\begin{equation}
F_{F}(\beta)=N_{F}f_{B}(\beta)-2N_{F}f_{B}(2\beta)
\end{equation}
Now it is easy to get the free energy for a field with bosonic and
fermionic degrees of freedom
\begin{equation}
F(\beta)=(N_{B}-N_{F})f_{B}(\beta)+N_{F}f_{S}(\beta)
\label{fields}
\end{equation}
where $f_{S}(\beta)=f_{B}(\beta)+f_{F}(\beta)$. If the
cosmological constant vanishes then the free energy can be written
\begin{equation}
F(\beta)=N_{F}f_{S}(\beta)=-2N_{F}
\int_{0}^{\infty}ds s^{-1-\frac{d}{2}}
\theta_{2}\left(0\left|\frac{2i\beta^{2}}{\pi s}\right.\right)
e^{-m^{2}s/2}
\end{equation}

Weighting the states in the scalar and vectorial
conjugacy classes as bosonic quantum fields and both spinorial classes
as
fermionic ones we obtain for the integrand of the $\beta-$dependent part
of the free energy
\begin{equation}
\chi(\tau,\bar{\tau})=
\left[\sum_{s}\Lambda_{s}(\tau,\bar{\tau})\right]
\theta_{3}^{'}\left(0\left|\frac{i\beta^{2}}{2\pi^{2}\tau_{2}}
\right.\right)-
\tau^{-\frac{d-2}{2}}_{2}
\frac{2\bar{\theta}_{2}^{4}}{\bar{\eta}^{12}}(z_{s}+z_{c})
\theta_{2}\left(0\left|\frac{2i\beta^{2}}{\pi^{2}\tau_{2}}\right.\right)
\end{equation}
Since $\chi(\tau,\bar{\tau})$ is invariant only under the Borel subgroup
generated by the transformation $T$ we have that the free energy is
obtained by integrating over the fundamental region of this
subgroup which is the strip $S=\{\tau_{1}+i\tau_{2} | \tau_{2}\geq 0,
-1/2\leq \tau_{1} < 1/2 \}$. Physically in the analog model we
integrate over the proper time from $0$ to $+\infty$ and at the same
time we have to impose the left-right level matching condition as a
Kronecker delta that finally becomes an integral over phases.
So we have that the $\beta-$dependent part of the free energy in the
$S-$representation is given by
\begin{equation}
F(\beta)=\int_{S}\frac{d^{2}\tau}{\tau_{2}^{2}}
\left\{
\left[\sum_{s}\Lambda_{s}(\tau,\bar{\tau})\right]
\theta_{3}^{'}\left(0\left|\frac{i\beta^{2}}{2\pi^{2}\tau_{2}}
\right.\right)-
\tau^{-\frac{d-2}{2}}_{2}
\frac{2\bar{\theta}_{2}^{4}}{\bar{\eta}^{12}}(z_{s}+z_{c})
\theta_{2}\left(0\left|\frac{2i\beta^{2}}{\pi^{2}\tau_{2}}
\right.\right)\right\}
\label{Simplfree}
\end{equation}
This is in perfect accordance with the field theoretical result
(\ref{fields}).

Our next task will be to go from the $S-$representation for the free
energy to the
$F-$ representation in which it is
expressed as an integral
over the fundamental region of the modular group of a function which is
invariant under the full modular group (cf. for example \cite{Osorio}).
To do this, we will make use of
the coset techniques developed in \cite{EalvaMar1} (see also
\cite{Moore}\cite{OBrien}). Let us separate (\ref{Simplfree}) into
two parts
\begin{equation}
F(\beta)=F_{1}(\beta)+F_{2}(\beta)
\end{equation}
with
\begin{eqnarray}
F_{1}(\beta)&=&\int_{S}\frac{d^{2}\tau}{\tau_{2}^{2}}
\left[\sum_{s}\Lambda(\tau,\bar{\tau})\right]
\theta_{3}^{'}\left(0\left|\frac{i\beta^{2}}{2\pi^{2}\tau_{2}}
\right.\right)
\\
F_{2}(\beta)&=&-\int_{S}\frac{d^{2}\tau}{\tau_{2}^{2}}
\tau^{-\frac{d-2}{2}}_{2}
\frac{2\bar{\theta}_{2}^{4}}{\bar{\eta}^{12}}(z_{s}+z_{c})
\theta_{2}\left(0\left|\frac{2i\beta^{2}}{\pi^{2}\tau_{2}}
\right.\right)
\end{eqnarray}

Since, as we have imposed above, $\sum_{s} \Lambda_{s}$ is a modular
invariant quantity, we can obtain the $F-$representation of
$F_{1}(\beta)$ almost inmediately
\begin{equation}
F_{1}(\beta)\rightarrow \int_{\cal F} \frac{d^{2}\tau}{\tau_{2}^{2}}
\left[ \sum_{s} \Lambda_{s}(\tau,\bar{\tau}) \right]
\theta^{'}\left[
\begin{array}{cc}
0 & 0 \\
0 & 0
\end{array}
\right](0|\Omega)
\label{scalvec}
\end{equation}
here the prime indicates again the absence of the zero-mode and
$\Omega={\bf \Omega}_{g}$ with $g=1$.

$F_{2}(\beta)$ can be rewritten in a modular invariant way by going from
the subgroup of discrete translations, $B$, to the congruence subgroup
$\Gamma_{0}(2) \subset \Gamma$ and finally to the full modular group
$\Gamma$ \cite{Mumford}\cite{Moore}\cite{EalvaMar1}. We get
\begin{eqnarray}
\lefteqn{F_{2}(\beta) \stackrel{B\rightarrow\Gamma_{0}(2)\rightarrow
\Gamma}{\longrightarrow}
\int_{\cal F} \frac{d^{2}\tau}{\tau_{2}^{2}}
\tau^{-\frac{d-2}{2}}_{2}\left\{
-\frac{2\bar{\theta}_{2}^{4}}{\bar{\eta}^{12}}(z_{s}+z_{c})
\theta \left[
\begin{array}{cc}
0  &  \frac{1}{2} \\
0  &  0
\end{array}
\right](0|4\Omega) \right.-} \nonumber \\
& & \left.\frac{2\bar{\theta}_{4}^{4}}{\bar{\eta}^{12}}(z_{v}-z_{o})
\theta \left[
\begin{array}{cc}
\frac{1}{2}  &  0 \\
0            &  0
\end{array}
\right](0|4\Omega)+
\frac{2\bar{\theta}_{3}^{4}}{\bar{\eta}^{12}}
(z_{v}+z_{o})
\theta \left[
\begin{array}{cc}
\frac{1}{2}  &  \frac{1}{2} \\
0            &  0
\end{array}
\right](0|4\Omega) \right\}
\label{spinor}
\end{eqnarray}
Collecting together (\ref{scalvec}) and (\ref{spinor}) we finally arrive
at the modular invariant expression of the Helmholtz free energy for a
general heterotic string
\begin{eqnarray}
\lefteqn{F(\beta)=
\int_{\cal F} \frac{d^{2}\tau}{\tau_{2}^{2}}
\left[ \sum_{s} \Lambda_{s} (\tau,\bar{\tau}) \right] \theta^{'} \left[
\begin{array}{cc}
0   &   0  \\
0   &   0
\end{array}
\right](0|\Omega)-
\int_{\cal F} \frac{d^{2}\tau}{\tau_{2}^{2}}
\tau^{-\frac{d-2}{2}}_{2} \left\{
\frac{2\bar{\theta}_{2}^{4}}{\bar{\eta}^{12}}(z_{s}+z_{c})
\times \right.} \mbox{ }  \nonumber  \\
  & & \hspace*{-1.4cm} \theta \left[
\begin{array}{cc}
0  &  \frac{1}{2} \\
0  &  0
\end{array}
\right](0|4\Omega) +
\frac{2\bar{\theta}^{4}_{4}}{\bar{\eta}^{12}}(z_{v}-z_{o})
\theta \left[
\begin{array}{cc}
\frac{1}{2}  &  0 \\
0            &  0
\end{array}
\right](0|4\Omega)-
\left.\frac{2\bar{\theta}_{3}^{4}}{\bar{\eta}^{12}}(z_{v}+z_{o})
\theta \left[
\begin{array}{cc}
\frac{1}{2}  &  \frac{1}{2} \\
0            &  0
\end{array}
\right](0|4\Omega) \right\}
\label{free}
\end{eqnarray}

After some computations one can show that (\ref{free}) is the same
we would have
obtained from the application of Atick and Witten's prescription as
written in (\ref{Freenergy}). Indeed, by  using (\ref{upper}) and with
the identifications
\begin{eqnarray}
\Lambda_{\frac{1}{2},0}(\tau,\bar{\tau}) &=&
\tau_{2}^{-\frac{d-2}{2}} \frac{\bar{\theta}_{2}^{4}}{\bar{\eta}^{12}}
\left(z_{s}(\tau,\bar{\tau})+z_{c}(\tau,\bar{\tau})\right)  \nonumber \\
\Lambda_{0,\frac{1}{2}}(\tau,\bar{\tau}) &=&
\tau_{2}^{-\frac{d-2}{2}} \frac{\bar{\theta}_{4}^{4}}{\bar{\eta}^{12}}
\left(z_{v}(\tau,\bar{\tau})-z_{o}(\tau,\bar{\tau})\right)   \\
\Lambda_{0,0}(\tau,\bar{\tau}) &=&
-\tau_{2}^{-\frac{d-2}{2}} \frac{\bar{\theta}_{3}^{4}}{\bar{\eta}^{12}}
\left(z_{v}(\tau,\bar{\tau})+z_{o}(\tau,\bar{\tau})\right)  \nonumber
\end{eqnarray}
(the contribution from the odd spin structure
$\Lambda_{\frac{1}{2},\frac{1}{2}}$ being zero) one gets
\begin{eqnarray}
F_{A-W}(\beta)= & &
\int_{\cal F} \frac{d^{2}\tau}{\tau_{2}^{2}} \left[\sum_{s}
\Lambda_{s}
\right]  \sum_{s_{1},s_{2}} \theta \left[
\begin{array}{cc}
  s_{1}  &   s_{2}  \\
    0    &    0
\end{array}
\right] (0|4\Omega) -
\int_{\cal F} \frac{d^{2}\tau}{\tau_{2}^{2}}
\left\{ 2\Lambda_{0,\frac{1}{2}} \theta \left[
\begin{array}{cc}
   \frac{1}{2}   &   0  \\
   0             &   0
\end{array}
\right]  (0|4\Omega) \right. \nonumber  \\
 &  & +  \left. 2\Lambda_{\frac{1}{2},0} \theta \left[
\begin{array}{cc}
   0   &   \frac{1}{2}       \\
   0   &     0
\end{array}
\right] (0|4\Omega)+
2\Lambda_{0,0}  \theta \left[
\begin{array}{cc}
   \frac{1}{2}   &      \frac{1}{2}       \\
   0             &        0
\end{array}
\right] (0|4\Omega) \right\}
\label{AW}
\end{eqnarray}
that is easily proven to be almost identical to (\ref{free}).
The difference is that in eq. (\ref{free}) the zero mode of the Riemann
theta function with vanishing characteristics is missing as the result
of removing the ultraviolet divergent vacuum energy in the original
analog model. The presence of this zero mode in (\ref{AW}) guarantees
that $F_{A-W}(\beta)$ obeys (\ref{duality}). In the case of (\ref{free})
we have that the relation which holds is
\begin{equation}
F(\beta)=\frac{\pi^{2}}{\beta^{2}}F\left(\frac{\pi^{2}}{\beta}\right)-
\left(1-\frac{\pi^{2}}{\beta^{2}}\right)\Lambda
\label{quasiduality}
\end{equation}
However, by simply adding
the cosmological constant to (\ref{free}) we get that $F(\beta)$
obeys (\ref{duality}). This corresponds to dropping the prime in the
Riemann theta function with null characteristics. In other words to have
duality as a property of the partition function we need to set the
result of the decompactification limit to the value of the cosmological
constant.

As a particular example, we are now going to apply the prescription
to get the free energy for
the cases in the family of heterotic
strings in two non-compact dimensions which
have been treated in \cite{MarMavaz}
(see also \cite{Moore} and \cite{GinspargVafa}).
The contribution associated with every $SO(10)$ conjugacy class is given
by \cite{MarMavaz}
\begin{eqnarray}
z_{o}(\tau,\bar{\tau}) &=&
\frac{1}{2}\bar{\theta}_{2}^{4}(\bar{\theta}_{3}^{4}
+\bar{\theta}_{4}^{4})[j(\tau)+r_{\Gamma}(1)-720]  \label{o}\\
z_{v} (\tau,\bar{\tau}) &=& \left[ \frac{1}{4}(\bar{\theta}_{2}^{8} +
\bar{\theta}_{3}^{8} +\bar{\theta}_{4}^{8})+\frac{1}{2}
\bar{\theta}_{3}^{4}\bar{\theta}_{4}^{4} \right]
[j(\tau)+r_{\Gamma}(1)-720] \label{v}\\
z_{s} (\tau,\bar{\tau})  &=& \left[\frac{1}{4}(\bar{\theta}_{2}^{8} +
\bar{\theta}_{3}^{8} +\bar{\theta}_{4}^{8})-\frac{1}{2}
\bar{\theta}_{3}^{4}\bar{\theta}_{4}^{4}\right]
[j(\tau)+r_{\Gamma}(1)-720] \label{s}\\
z_{c}(\tau,\bar{\tau}) &=& \frac{1}{2}\bar{\theta}_{2}^{4}
(\bar{\theta}_{3}^{4} -\bar{\theta}_{4}^{4})[j(\tau)+r_{\Gamma}(1)-720]
\label{c}
\end{eqnarray}
where $j(\tau)$ is the modular invariant function and $r_{\Gamma}(1)$ is
the number of lattice vectors with (length)$^{2}=2$ which parametrizes
the $24$ selfdual Niemeier lattices among which we choose one to
compactify the left moving coordinates.

By using (\ref{free}) we obtain the free energy per unit volume for
the family of heterotic strings described above as
\begin{eqnarray}
\lefteqn{F(\beta)=-48\int_{\cal F} \frac{d^{2}\tau}{\tau^{2}_{2}}
[j(\tau)+r_{\Gamma}(1)-720] \theta \left[
\begin{array}{cc}
  0  &  0  \\
  0  &  0
\end{array}
\right] (0|\Omega)-
\int_{\cal F} \frac{d^{2}\tau}{\tau^{2}_{2}}
[j(\tau)+r_{\Gamma}(1)-720]}  \nonumber \\
 & & \times \left\{
\frac{2\bar{\theta}_{2}^{12}}{\bar{\eta}^{12}}
\theta \left[
\begin{array}{cc}
  0   &   \frac{1}{2}  \\
  0   &    0
\end{array}
\right](0|4\Omega) +
\frac{2\bar{\theta}_{4}^{12}}{\bar{\eta}^{12}}
\theta \left[
\begin{array}{cc}
 \frac{1}{2} &  0     \\
     0       &  0
\end{array}
\right](0|4\Omega) -
\frac{2\bar{\theta}_{3}^{12}}{\bar{\eta}^{12}}
\theta \left[
\begin{array}{cc}
  \frac{1}{2}  &  \frac{1}{2}  \\
      0        &      0
\end{array}
\right](0|4\Omega) \right\}
\label{twodim}
\end{eqnarray}
where we have set all the constants in front of the integrals to
one.

\section{Bosonic and fermionic degrees of freedom in heterotic strings}

In the analog model the presence of the Hagedorn length is seen as
resulting from the exponential growth of the degeneracy of the states in
a mass level when the mass goes to infinity. So when this phenomenon
happens it does not disappear by increasing the temperature; in fact it
worsens. The strange feature implied by duality in the modular invariant
expression is that below $\beta_{H}^{*}$ the free energy is finite and
then the analog model interpretation does not hold. One may ask oneself
what happens to the density of states we could associate with each of
the two statistics. The standard in quantum field theory is that the
free energy diverges as $-\beta^{-d}$ when $\beta$ goes to zero
(see \cite{Osorio} and references therein and also \cite{AtickWitten}).
For a system of bosons and fermions the total
free energy behaves in this limit as
\begin{equation}
F_{T}(\beta) \sim -\left[N_{F}(1-2^{1-d})+N_{B}\right]\beta^{-d}
\end{equation}
In particular for a supersymmetric theory
\begin{equation}
F_{T}(\beta) \sim -N_{T}(1-2^{-d})\beta^{-d}
\end{equation}
where $N_{T}=N_{F}+N_{B}=2N_{B}$. Thus the combined action of bosons and
fermions never results in a cancelation between the two types of degrees
of freedom.

In this section we will show that it is possible to obtain separated
modular
invariant expressions for the free energy of each of the the two species
of degrees of freedom in the heterotic string: bosonic
and fermionic ones. We will see that for each case the corresponding
free energy is infinite from $\beta_{H}$ to $\beta=0$. It will be only
after combining the two contributions that the strange high temperature
phase appears as though there were tachyons in the fermionic sector
having and preserving the fermionic character at high temperature.

To get the modular invariant expression we start using the analog model
as follows from the prescription given in the last section to get
\begin{equation}
F_{B}(\beta)=\int_{\cal S} \frac{d^{2}\tau}{\tau_{2}^{2}}
\tau_{2}^{-\frac{d-2}{2}}\left[
\frac{\bar{\theta}_{4}^{4}}{\bar{\eta}^{12}}(z_{v}-z_{o})-
\frac{\bar{\theta}_{3}^{4}}{\bar{\eta}^{12}}(z_{v}+z_{o})\right]
\theta^{'}_{3}\left(0\left|\frac{i\beta^{2}}{2\pi^{2}\tau_{2}}
\right.\right)
\label{Sbos}
\end{equation}
for bosons and
\begin{equation}
F_{F}(\beta)=\int_{S}\frac{d^{2}\tau}{\tau_{2}^{2}}
\tau_{2}^{-\frac{d-2}{2}}\frac{\bar{\theta}_{2}^{4}}
{\bar{\eta}^{12}}(z_{s}+z_{c})
\theta^{'}_{4}\left(0\left|\frac{i\beta^{2}}{2\pi^{2}\tau_{2}}
\right.\right)
\label{Sferm}
\end{equation}
for fermions. Now, the goal is to obtain from (\ref{Sbos}) and
(\ref{Sferm})
a modular invariant result. It is not straightforward to get manageable
expressions. By manageable we mean written in terms of our useful
Riemann theta functions.
The reason is that
the behavior of the Jacobi theta functions under the modular group
transformations do not correspond to
that of the terms that multiply each of them. To solve this
problem we use the easily proven relations
\begin{eqnarray}
\theta_{3}^{'}\left(0\left|\frac{i\beta^{2}}{2\pi^{2}\tau_{2}}\right.
\right)
&=&\theta_{3}^{'}\left(0\left|\frac{i(2\beta)^{2}}{2\pi^{2}\tau_{2}}
\right.
\right) +\theta_{2}\left(0\left|\frac{i(2\beta)^{2}}{2\pi^{2}\tau_{2}}
\right.\right) \\
\theta_{4}^{'}\left(0\left|\frac{i\beta^{2}}{2\pi^{2}\tau_{2}}
\right.\right)
&=&\theta_{3}^{'}\left(0\left|\frac{i(2\beta)^{2}}{2\pi^{2}\tau_{2}}
\right.\right)
-\theta_{2}\left(0\left|\frac{i(2\beta)^{2}}{2\pi^{2}\tau_{2}}
\right.\right)
\end{eqnarray}
to get the following expansions
\begin{equation}
\theta_{3}^{'}\left(0\left|\frac{i\beta^{2}}{2\pi^{2}\tau_{2}}
\right.\right) =
\theta_{3}^{'}\left(0\left|\frac{i(2^{N}\beta)^{2}}{2\pi^{2}\tau_{2}}
\right.\right) + \sum_{k=1}^{N}
\theta_{2}\left(0\left|\frac{i(2^{k}\beta)^{2}}{2\pi^{2}\tau_{2}}
\right.\right)
\end{equation}
and
\begin{equation}
\theta_{4}^{'}\left(0\left|\frac{i\beta^{2}}{2\pi^{2}\tau_{2}}
\right.\right) =
\theta_{3}^{'}\left(0\left|\frac{i(2^{N}\beta)^{2}}{2\pi^{2}\tau_{2}}
\right.\right) + \sum_{k=2}^{N}
\theta_{2}\left(0\left|\frac{i(2^{k}\beta)^{2}}{2\pi^{2}\tau_{2}}
\right.\right)-
\theta_{2}\left(0\left|\frac{i(2\beta)^{2}}{2\pi^{2}\tau_{2}}
\right.\right)
\end{equation}
We are now prepared to consider the case of the bosons; their free
energy can be written as
\begin{equation}
F_{B}(\beta)=F_{B}(2^{N}\beta)+\sum_{k=1}^{N}\int_{\cal S}
\frac{d^{2}\tau}{\tau_{2}^{2}} \tau_{2}^{-\frac{d-2}{2}}\left[
\frac{\bar{\theta}_{4}^{4}}{\bar{\eta}^{12}}(z_{v}-z_{o})-
\frac{\bar{\theta}_{3}^{4}}{\bar{\eta}^{12}}(z_{v}+z_{o})\right]
\theta_{2}\left(0\left|\frac{i(2^{k}\beta)^{2}}{2\pi^{2}\tau_{2}}
\right.\right)
\end{equation}
We can easily go from the Borel subgroup to $\Gamma_{0}(2)$ and
from this to the full modular group. Following this procedure and
taking finally the limit $N \rightarrow \infty$ (what is perfectly
justified for $\beta \neq 0$),we arrive at
\begin{eqnarray}
 F_{B}(\beta)=\int_{\cal F}
\frac{d^{2}\tau}{\tau_{2}^{2}}\left[\sum_{s} \Lambda_{s} \right]
\theta^{'} \left[
\begin{array}{cc}
 0  &  0  \\
 0  &  0
\end{array}
\right] (0|\Omega) + \hspace*{4cm} \nonumber \\
\hspace*{.4cm} \frac{1}{2} \sum_{k=0}^{\infty} \left\{
F(2^{k}\beta)-\int_{\cal F}
\frac{d^{2}\tau}{\tau_{2}^{2}}
\left[\sum_{s} \Lambda_{s} \right] \theta^{'}\left[
\begin{array}{cc}

 0  &  0  \\
 0  &  0
\end{array}
\right](0|2^{2k}\Omega)  \right\}
\label{boson}
\end{eqnarray}
where $F(\beta)$ concides with the total free energy given by
(\ref{free}).
For the fermionic degrees of freedom we operate along the same lines;
its contribution to the free energy can be written
as
\begin{eqnarray}
F_{F}(\beta)=
F(\beta)-\int_{\cal F}
\frac{d^{2}\tau}{\tau_{2}^{2}}\left[\sum_{s} \Lambda_{s} \right]
\theta^{'} \left[
\begin{array}{cc}
 0  &  0  \\
 0  &  0
\end{array}
\right] (0|\Omega) -
\frac{1}{2} \sum_{k=0}^{\infty} \left\{
F(2^{k}\beta)- \right.\nonumber \\
\left.\int_{\cal F} \frac{d^{2}\tau}{\tau_{2}^{2}}
\left[\sum_{s} \Lambda_{s} \right] \theta^{'}\left[
\begin{array}{cc}

 0  &  0  \\
 0  &  0
\end{array}
\right](0|2^{2k}\Omega)  \right\}
\label{fermion}
\end{eqnarray}
which is indeed equal to $F(\beta)-F_{B}(\beta)$.

For the case of supersymmetric heterotic strings (\ref{boson}) and
(\ref{fermion}) drastically simplify. The bosonic contribution to the
free energy is simply
\begin{equation}
F_{B}(\beta)=\frac{1}{2}\sum_{k=0}^{\infty} F(2^{k}\beta)
\label{hetbos}
\end{equation}
The singularity structure of (\ref{hetbos}) is given by an infinite set
$\Sigma$ of values of $\beta$ labeled by the integer $k$ running in the
sum:
$\Sigma= \{\beta_{k}\;|\;\beta_{k}=2^{-k}\beta_{H}\}$ where $\beta_{H}$
is the Hagedorn length of the heterotic string. $\beta-$duality for
each term of the sum (\ref{hetbos}) implies a second set of
lengths
$\Sigma^{*}=\{\beta_{k}^{*}\;|\;\beta_{k}^{*}=\pi^{2}/\beta_{k}\}$.
Then we find that for each interval
$(\beta^{*}_{k},\beta_{k})$ at least one of the terms of the sum in
(\ref{hetbos}) diverges. It is also easy to show that this family of
intervals overlap, so $F_{B}(\beta)$ diverges for any $\beta<\beta_{H}$.
For the case of fermions the situation is exactly the same. We then see
that duality is a property of the total free energy and, in a
first sight, results from the mutual cancelation between the divergences
of the bosonic and fermionic parts so as to leave only an interval of
divergence given by $(\beta_{H}^{*},\beta_{H})$.

However, looking carefully, duality invariance is already sowed in both
contributions. Let
us take the limit $\beta \rightarrow 0^{+}$ in $F_{B(F)}(\beta)$. By
duality symmetry for each
term in the sum over $k$, $F_{B(F)}(\beta)$ will go to zero, and then we
have that no bosonic
(fermionic) degree of freedom survives \cite{Osorio}.

For the general case, computing the $\beta \rightarrow 0^{+}$ limit
of $F_{B}(\beta)$ and
$F_{F}(\beta)$, we get the paradoxical results
\begin{eqnarray}
F_{B}(\beta) \sim \frac{\pi^{2}}{\beta^{2}}\left( \frac{4}{3} \Lambda
\right) \\
F_{F}(\beta) \sim -\frac{\pi^{2}}{\beta^{2}}\left( \frac{1}{3} \Lambda
\right) \end{eqnarray}
The solution of this paradox is that the equivalence between
the corresponding analog models and the modular invariant expressions
is broken.
A related question is that of the value of $\beta$ (i. e., the energy
scale)
at which the intruder fermionic (bosonic) degrees of freedom corrupt
the bosonic (fermionic) free energy. We postpone this problem until
the next section.

\section{The singularities of the free energy of a heterotic string}

In quantum field theory the equivalence between the free energy of a
quantum field and the contribution to the vacuum energy --cosmological
constant-- of the same field with an euclidean time is a very
well known fact (cf. e.g. \cite{Osorio} and references therein).
In the case of strings
\cite{Polchinski}\cite{EalvaMar1}\cite{AtickWitten} this
relationship is not direct. Actually, it depends on whether the string
can be described by its field content. Until now the answer has not been
too precise; in general, the fact that any string model has a Hagedorn
temperature has been used to break any relationship at higher
temperatures. From the work done in \cite{KutasovSeiberg}\cite{MarMavaz}
it has been concluded that in some compactifications this equivalence is
broken by a different kind of singularities of the
partition function.

The strategy is to find the relationship between the modular invariant
free energy gotten by using the coset technique and the cosmological
constant of a zero temperature heterotic theory. Then we will study the
analytic behavior of the free energy through the mass formulae of the
corresponding zero temperature theory. After that we will see that,
at the Hagedorn temperature, the
analog model presents the same divergent behavior as the modular
invariant extension. The breakdown point
will be the self-dual radius
below which we claim that the equivalence between the modular invariant
result and the analog model does not hold any more.

In what follows we are going to treat this problem for
the general
free energy of a heterotic string. To prove that the manifestly modular
invariant free energy corresponds in every case to the cosmological
constant of a heterotic string with an euclidean time it is useful to
realize that, after some Riemann theta-functions gymnastics, the
integrand of (\ref{free}) can be written (we now drop the
multiplicative factor $\tau_{2}^{-(d-2)/2}$)
\begin{eqnarray}
\chi(\tau,\bar{\tau})=
-\frac{\bar{\theta}^{4}_{3}-\bar{\theta}^{4}_{4}}{\bar{\eta}^{12}}
\left\{ z_{o} \left( \theta \left[
\begin{array}{cc}
 \frac{1}{2}  &  0 \\
  0           &  0
\end{array}\right] -
\theta\left[
\begin{array}{cc}
 \frac{1}{2}  &  \frac{1}{2} \\
  0           &  0
\end{array}\right] \right) +
z_{v}\left(\theta \left[
\begin{array}{cc}
  0           &  0 \\
  0           &  0
\end{array}\right] +
\theta\left[
\begin{array}{cc}
  0           &  \frac{1}{2} \\
  0           &  0
\end{array}
\right] \right) \right\} \nonumber \\
-\frac{\bar{\theta}^{4}_{3}+\bar{\theta}^{4}_{4}}{\bar{\eta}^{12}}
\left\{z_{o}\left(\theta \left[
\begin{array}{cc}
  0           &  0 \\
  0           &  0
\end{array}\right] +
\theta \left[
\begin{array}{cc}
  0           &  \frac{1}{2} \\
  0           &  0
\end{array}\right] \right) +
z_{v}\left(\theta \left[
\begin{array}{cc}
  \frac{1}{2} &  0 \\
  0           &  0
\end{array}\right] -
\theta\left[
\begin{array}{cc}
  \frac{1}{2} &  \frac{1}{2} \\
  0           &  0
\end{array}
\right] \right) \right\} \nonumber \\
+\frac{\bar{\theta}^{4}_{2}}{\bar{\eta}^{12}}
\left\{z_{s}\left(\theta \left[
\begin{array}{cc}
  0           &  0 \\
  0           &  0
\end{array}\right] -
\theta\left[
\begin{array}{cc}
  0           &  \frac{1}{2} \\
  0           &  0
\end{array}\right] \right) +
z_{c}\left(\theta  \left[
\begin{array}{cc}
  \frac{1}{2} &  0 \\
  0           &  0
\end{array}\right] +
\theta\left[
\begin{array}{cc}
  \frac{1}{2} &  \frac{1}{2} \\
  0           &  0
\end{array}
\right] \right) \right\} \nonumber \\
+\frac{\bar{\theta}^{4}_{2}}{\bar{\eta}^{12}}
\left\{z_{s}\left(\theta \left[
\begin{array}{cc}
  \frac{1}{2} &  0 \\
  0           &  0
\end{array}\right] +
\theta\left[
\begin{array}{cc}
  \frac{1}{2} &  \frac{1}{2} \\
  0           &  0
\end{array}\right] \right) +
z_{c}\left(\theta \left[
\begin{array}{cc}
  0           &  0 \\
  0           &  0
\end{array}\right] -
\theta\left[
\begin{array}{cc}
  0           &  \frac{1}{2} \\
  0           &  0
\end{array}
\right] \right) \right\}
\label{compact}
\end{eqnarray}
where all the Riemann theta functions are evaluated at $(0|4\Omega)$.
With this form of the free energy it is also easy to read back the
mass formulae and constraints for each sector in the corresponding
theory at zero temperature. The general structure of eq. (\ref{compact})
can be sketched as
$(v,\Gamma_{v})+(o,\Gamma_{o})+(s,\Gamma_{s})+(c,\Gamma_{c})$
where $v,o,s,c$ are the conjugacy classes of $SO(10)$ and $\Gamma_{i}$
with $i=v,o,s,c$ are sets of vectors (that in general do not close under
addition) defined in the following way: corresponding to each of the two
terms multiplying the contribution of the conjugacy classes of $SO(10)$
we define two sets, $\Gamma_{i,1}$ and $\Gamma_{i,2}$ such
that $\Gamma_{i}=\Gamma_{i,1} \bigcup \Gamma_{i,2}$. Now, for each
class, we can characterize these sets as follows
\begin{eqnarray}
\Gamma_{i,1}&=&\left\{ ({\bf p}_{L},{\bf p}_{R}) | {\bf p}_{L,R}=
\left({\vec v}_{L,R},p_{L,R}(\beta) \right) \right\} \nonumber \\
\Gamma_{i,2}&=&\left\{ ({\bf p}_{L},{\bf p}_{R}) | {\bf p}_{L,R}=
\left({\vec w}_{L,R},p_{L,R}(\beta)+\delta_{L,R}(\beta)\right) \right\}
\end{eqnarray}
Here the first entry corresponds to the vectors given the generalized
theta-functions containted in $z_{i}$ which may eventually depend on a
radius of
compactification, although there are always components which correspond
to internal compact dimensions at fixed radii, at least $16$ dimensions
in the left-moving sector.
The second one is a
function of $\beta$ corresponding to the momentum in the euclidean time
and $\delta_{L,R}(\beta)$ is a shift.

The mass formulae and constrains for the four bosonic
sectors in (\ref{compact}) are
\begin{eqnarray}
\frac{1}{4}m^{2}_{i,1}=N_{R}^{i}+N_{L}+\frac{1}{2}p_{R,i+1}^{2}
+\frac{1}{2}p_{L,i+1}^{2}+\frac{\beta^{2}}{4\pi^{2}}(2m+1)^{2}+
\frac{\pi^{2}}{4\beta^{2}}(2n+1)^{2}-\frac{3}{2}  \label{mass1} \\
N_{L}-N_{R}^{i}+\frac{2m+1}{2} (2n+1)+\frac{1}{2}p_{L,i+1}^{2}-
\frac{1}{2}p_{R,i+1}^{2}-\frac{1}{2}=0 \label{mass2} \\
\frac{1}{4}m^{2}_{i,2}=N_{R}^{i}+N_{L}+\frac{1}{2}p_{R,i+1}^{2}
+\frac{1}{2}p_{L,i+1}^{2}+\frac{\beta^{2}}{4\pi^{2}}(2m)^{2}+
\frac{\pi^{2}}{4\beta^{2}}(2n)^{2}-\frac{3}{2} \\
N_{L}-N_{R}^{i}+2mn+\frac{1}{2}p_{L,i+1}^{2}-
\frac{1}{2}p_{R,i+1}^{2}-\frac{1}{2}=0
\end{eqnarray}
where $i=0,1$ ({\it v} $\equiv 0$ (mod 2), {\it o} $\equiv 1$ (mod 2));
$N_{L}$ is a positive integer and $N_{R}^{o} \in {\bf Z^{+}}$,
$N_{R}^{v} \in {\bf Z^{+}}+1/2$; again $p_{R(L),i}$ denotes de vectors
associated with each $z_{i}$.
For the four fermionic sectors we have
\begin{eqnarray}
\frac{1}{4}m^{2}_{j,1}=N_{R}^{j}+N_{L}+\frac{1}{2}p_{R,j}^{2}
+\frac{1}{2}p_{L,j}^{2}+\frac{\beta^{2}}{4\pi^{2}}(2m)^{2}+
\frac{\pi^{2}}{4\beta^{2}}(2n+1)^{2}-1\\
N_{L}-N_{R}^{j}+m(2n+1)+\frac{1}{2}p_{L,j}^{2}-
\frac{1}{2}p_{R,j}^{2}-1=0 \\
\frac{1}{4}m^{2}_{j,2}=N_{R}^{j}+N_{L}+\frac{1}{2}p_{R,i}^{2}
+\frac{1}{2}p_{L,j}^{2}+\frac{\beta^{2}}{4\pi^{2}}(2m+1)^{2}+
\frac{\pi^{2}}{4\beta^{2}}(2n)^{2}-1 \\
N_{L}-N_{R}^{j}+(2m+1)n+\frac{1}{2}p_{L,j}^{2}-
\frac{1}{2}p_{R,j}^{2}-1=0
\end{eqnarray}
where now $j=0,1$ ({\it s} $\equiv 0$ (mod 2), {\it c} $\equiv 1$ (mod
2)); and
$N_{R}^{j},N_{L}^{j}$ are positive integers.
{}From these formulae one can show that the Hagedorn length is the
same for every heterotic string (cf. \cite{Hagedorn}); it comes from the
sector associated with the scalar conjugacy class and $z_{v}$ with the
following set of quantum numbers: $N_{L}=N_{R}=0$,
$m=n=0$ or $m=n=-1$ and $p_{L,v}^{2}=p_{R,v}^{2}=0$.

The existence of a vanishing value of the momentum in the compact
dimensions associated with the vectorial conjugacy class in the theory
at zero temperature results from the induced $U(1)$ Kaluza-Klein
bosons.
That these bosons must appear glued to this class is the result of
modular
invariance because $T(z_{v})=z_{v}$ and then if $(p_{L,v};p_{R,v})$ is
a vector we must have $p_{L,v}^{2}-p_{R,v}^{2}\equiv 0 \mbox{(mod 2)}$.
Since $z_{v}$ in general receives contributions from
vectors $(p_{L,v};p_{R,v})=(\tilde {p}_{L,v},p_{L,v}(R);\hat {p}_{R,v},
p_{R,v}(R))$ where $\tilde {p}$ and $\hat{p}$ are momenta corresponding
to fixed size dimensions (at least sixteen dimensions in the left
sector), and $p_{L,v}(R)$, $p_{R,v}(R)$ are left and right-momenta
depending on the value of the radii; it would appear that there is a way
of getting $T(z_{v})=z_{v}$ by having  $p_{L,v}^{2}-p_{R,v}^{2}\equiv 1
\mbox{(mod 2)}$ and $\hat{p}^{2}-\tilde{p}^{2}\equiv 1 \mbox{(mod
2)}$. This is impossible because this would imply that in the limit in
which the radii of compactification go to infinity $z_{v}$ goes to
zero and then the corresponding theory in the decompactification limit
could not be modular invariant because of the absence of the
contribution of the vectorial conjugacy class. On the contrary,
$T(z_{o})=-z_{o}$ implies that
$p_{L,o}^{2}-p_{R,o}^{2}\equiv 1 \mbox{(mod 2)}$.

The Hagedorn temperature, which corresponds to the length of the
euclidean time $\beta_{H}=\pi(\sqrt{2}+1)$, and its dual
($1/T_{H}^{*}=\beta_{H}^{*}=\pi(\sqrt{2}-1)$) are singularities at which
the
free energy diverges respectively to the left and to the right in a plot
$F(\beta)$ vs. $\beta$.

In the same sector we can also find that for $\{ N_{L}=1$, $N_{R}=0$,
$m=0, n=-1$ or $m=-1, n=0$ and $p_{L,v}^{2}=p_{R,v}^{2}=0\}$ or
$\{ N_{L}=N_{R}=0$, $m=0, n=-1$ or $m=-1, n=0$ and $p_{L,v}^{2}=2$,
$p_{R,v}^{2}=0$\}, we have
another
critical length at $\beta=\pi$, the self-dual point of the free
energy. This value of $\beta$ gives $m^{2}=0$ but such that
$m^{2}>0$ for every $\beta \neq \pi$.

Both lengths are generic for every heterotic string, supersymmetric
or not, as long as we have Kaluza-Klein $U(1)$ bosons. Of course there
may be more critical lengths as the Hagedorn one
and their duals that depend on $R$ \cite{Tan} and consequently are
not generic. Furthermore, there is an intermediate case. When there is a
vector of (length)$^2=0$ in one of the spinorial representations
\footnote{This suffices since
duality symmetry interchanges both spinorial conjugacy classes and then
both chiralities must have a null vector in each of the associated sets
of vectors.} we have a class of heterotic string theories for which two
more critical lengths, $\beta=\pi\sqrt{2}$ and its dual
$\beta^{*}=\pi/\sqrt{2}$, can be found
independently of the radii and such that $m^2>0$ for every  $\beta \neq
\pi\sqrt{2},\pi/\sqrt{2}$.

For example, in the first case, in which only Hagedorn and the self-dual
point are
generic temperatures, we find the non-supersymmetric $O(16)\times O(16)$
\cite{LAG}\cite{NairShapereWil} and in general any heterotic theory for
which $z_{s}$ and $z_{c}$ do not receive a contribution of vectors with
(length)$^{2}=0$. In particular the model presented in section 2 whose
free energy is given by (\ref{twodim}) does not present these
singularities since the corresponding generalized theta functions
associated with $z_{s}$ and $z_{c}$
start with powers of $q$ higher than zero (see (\ref{s}) and (\ref{c})).
On the other hand for the old heterotic string
\cite{Pricentonquartet} we have
the opposite situation, because the $E_{8}\times E_{8}$ lattice, which
is common to all conjugacy classes, do have a vector with $p_{L}^{2}=0$.

When the term producing the Hagedorn
temperature is substracted from the free energy, the remaining would-be
renormalized free energy presents a singularity at the self-dual point.
The term contributing to the free energy associated with this
singularity is
\begin{eqnarray}
I(\beta)=-2\frac{\pi\sqrt{2}}{\beta} \int_{1}^{\infty} d\tau_{2}
\tau_{2}^{-\frac{d+1}{2}}
\exp{\left[-\frac{\pi}{2}\tau_{2}\left(\frac{\beta^{2}-\beta_{0}^{2}}
{\beta \beta_{0}} \right)^{2}\right]} = \nonumber\\
\frac{2\pi\sqrt{2}}{\beta}\left[\pi\left(\frac{\beta^{2}-\beta_{0}^{2}}
{\beta\beta_{0}}\right)^{2}\right]^{\frac{d-1}{2}}
\Gamma\left[\frac{1-d}{2},\frac{\pi}{2}\left(\frac{\beta^{2}
-\beta_{0}^{2}} {\beta\beta_{0}}\right)^{2}\right]
\label{singular}
\end{eqnarray}
where $\beta_{0}=\pi$. Whenever $d \geq 2$ and even $I(\beta)$ is
finite for any value of $\beta$; in particular, at the
corresponding singular point it has a finite jump
in the $d-1$ derivative. For $d$ odd it suffers from a
logarithmic
singularity in the $d-1$ derivative at $\beta_{0}$.
When $d=0,1$ the
contribution of $I(\beta)$ to the free energy is divergent at
$\beta=\beta_{0}$ but finite at any other point.
It is worth noticing that $\beta_{0}^{-1}I(\beta)$ is always invariant
under the exchange $\beta \leftrightarrow \beta_{0}$. This implies that
near $\beta_{0}$, $\beta_{0}^{-1}I(\beta)$ is a function of
$|\beta-\beta_{0}|$.

When the singularities at $\pi\sqrt{2}$ and $\pi/\sqrt{2}$ appear
the associated terms contribute to the free energy as in
(\ref{singular}) with the opposite sign and a factor $2$
multiplying the argument of the exponential function in the integrand.
Finally we should note that all these terms must be multiplied by
integral degeneracy factors. For example for the ten dimensional
supersymmetric heterotic string we would have a factor given by
$24+2\times240=504$ multiplying $I(\beta)$ with $\beta_{0}=\pi$.

\section{Conclusions}

Let us go back in time to reference \cite{MarMavaz} where the
dependence
of the partition function on the compactification scale was studied for
a particular class of two dimensional heterotic strings. There we showed
that no critical radius as the Hagedorn one appears. There is
of course a singularity, but of a quite different kind.
Furthermore, it is easy to show that the origin of this soft
singularity is the unstability of the vacuum represented as
\begin{eqnarray}
-RM(R)& = &-R\int_{\cal
F}\frac{d^{2}\tau}{\tau_{2}^{2}}J(\tau){\sum_{m,n \in {\bf Z}}}^{'}
\exp{\left(-\frac{2\pi
R^{2}}{\tau_{2}}|m\tau+n|^{2}\right)}= \nonumber \\
& &  \mbox{} -4\pi R \left(1-\frac{1}{2R^{2}}\right)+
4\pi R \left|1-\frac{1}{2R^{2}}\right|
\label{vacuum}
\end{eqnarray}
Below the Planck length scale this way of representing nothingness
breaks down. A non-vanishing contribution appears which corresponds to a
massless field in two dimensions with positive Helmholtz free energy. In
fact $-48(RM(R)+\frac{4\pi}{R})$ equals (choosing adequate units to
eliminate $\pi$'s and
other constant factors) the partition function in ${\bf R}\times S^{1}$
of the non-sypersymmetric heterotic string described in
\cite{Harvey}\cite{Moore} which has Atkin-Lehner symmetry \cite{Moore}.
The term
$-RM(R)$ acts effectively as though it represented the contribution of a
ghost field for the net number of bosonic
degrees of freedom of the theory in ${\bf R}^{2}$ so as to destroy them.
This phenomenon is generic for this kind of trivial compactifications
in which the solitonic contribution is a common factor for all the
conjugacy classes; the bigger $d$ is the softer the singularity.
One would like to know whether this phenomenon at the selfdual
radius appears for other kind of compactifications, and if so whether it
is related to ``heteroticity'', i.e., the property of being a hybrid of
a bosonic and a fermionic string.

What we have shown in the present work is that by substracting from
the free energy a term $G(\beta)$ which gives the infinite background
for $\beta_{H}^{*}<\beta<\beta_{H}$ associated
with the Hagedorn singularity (whose contribution is negligible when
$\beta<\beta_{H}^{*}$) we can see a completely analogous phenomenon for
a big family of heterotic strings gotten by compactifying
non-trivialy one of the originally uncompact dimensions.
This non-trivial compactification has been dictated by the process of
getting the Helmholtz free energy for any heterotic string,
supersymmetric or not.
Looking at (\ref{singular}) when $d=2$ we see that at the
self-dual point (in this case, $\sqrt{\alpha^{'}}/2$) there is a finite
jump in the first derivative with a sign which is only consistent with a
loss of degrees of freedom. In particular for the supersymmetric
heterotic string the selfdual point corresponds to a cusp pointing to
the minus infinity direction in the graph of
$F(\beta)-G(\beta)$ vs. $\beta$. When $d>2$ this cusp softens to give a
cup. In the mass formulae the presence of this critical
compactification length depends upon the existence of the $U(1)$
Kaluza-Klein bosons associated with the Cartan subalgebra of the gauge
group.

Regarding the problem from a thermodynamical
point of view  we have shown that the theory
presented
in \cite{MarMavaz} whose free energy is given by
(\ref{twodim})
is an example of a theory with a thermal free energy which is dual and
at the same
time is a monotonous increasing function of $\beta$ in the would-be high
temperature phase. The complementary situation can be exemplified by the
$O(16)\times O(16)$ theory whose cosmological constant is positive
\cite{LAG}.

Another related point is that of the associated density of states
as a function of the compactification scale.
The main issue is to try to
get the density of states by inverse Laplace transforming the partition
function as a function of $R$. To do that we need to know the analytic
continuation of $Z(R)$ to the complex $R-$plane, but how
can we perform
the analytic continuation of a real function involving an absolute
value?
For example in the case of the Atkin-Lehner symmetric theory, if we try
to naively analytically
continue $-RM(R)$ to the $R$ complex plane by directly substituting
into the solitonic sum the real $R$ variable by a complex $R$, we would
have
obtained a complex function having, besides two singular points at
$\pm \sqrt{\alpha^{'}}$, a dense countable set of singularities
located over the imaginary axis (in fact all of the same type).
In the same way, if we substitute $R$ by $iR$ into the last part
of (\ref{vacuum}) we get that on the imaginary axis there
are only three singular points located at $0$ and
$\pm i\sqrt{\alpha^{'}}$. The lesson to be learned is that, in
general,
analytically continuing each term in the series in the integrand is not
equivalent to the continuation of the integrated result.
May be we can relax this requirement and demand only some kind of
procedure to perform the inverse Laplace transform without really
writing down the analytical continuation.

Finally it would also be interesting to know how string field theory
\cite{Kugo} might deal with these singularities when one looks at the
dependence of the
action on the parameters of the background target; in particular
regarding the
structure of the possible unitary transformation relating by
duality different backgrounds of the same class. Of course,
a prerequisite would be having a heterotic superstring field theory.

\section*{Acknowledgements}

We thank E. \'Alvarez for carefully reading the manuscript and
suggesting useful modifications. This work has been partially supported
by the CICyT proyect No. AEN-90-0272.

\end{document}